\documentclass[pre,aps,twocolumn,showpacs]{revtex4}
\input epsf

\usepackage{amssymb}
\usepackage{graphicx}
\usepackage{amsmath}
\usepackage{latexsym}
\usepackage{verbatim}

\begin{document}

\title{Effects of substrate network topologies on competition dynamics}

\author{Sang Hoon Lee}
 \email{lshlj@stat.kaist.ac.kr}
\author{Hawoong Jeong}
 \email{hjeong@kaist.ac.kr}
\affiliation {Department of Physics, Korea Advanced Institute of
Science and Technology, Daejeon 305-701, Korea}

\date{\today}

\begin{abstract}
We study a competition dynamics, based on the minority game, endowed
with various substrate network structures. We observe the effects of
the network topologies by investigating the volatility of the system
and the structure of follower networks. The topology of substrate
structures significantly influences the system efficiency
represented by the volatility and such substrate networks are shown
to amplify the herding effect and cause inefficiency in most cases.
The follower networks emerging from the leadership structure show a
power-law incoming degree distribution. This study shows the
emergence of scale-free structures of leadership in the minority
game and the effects of the interaction among players on the
networked version of the game.
\end{abstract}

\pacs{02.50.Le, 87.23.Ge, 89.65.Gh, 89.75.Hc}

\maketitle

\section{introduction}
The minority game (MG) captures the features of bounded rationality
and inductive reasoning, which are the radically different viewpoint
from that of the traditional economics, and has become a
representative model of competition
dynamics~\cite{Arthur1994,Challet1997,MGweb,ChalletBook,CoolenBook}.
MG has been intensively studied by adopting the agent-based modeling
and statistical mechanical approaches analogous to the spin glass
theory~\cite{Savit1999,Johnson1999,Hart2001,Cavagna1999a,Challet2000,
Challet1999,Cavagna1999,Challet2000b,Marsili2001a}. MG is a repeated
game where players choose one out of two alternatives at each time
step, which separates the entire group into two subgroups according
to the players' choice. The essential rule of MG is that those who
happen to be in the minority subgroup with fewer members win each
round of the game. Players watch the $m$ (memory) last winning
choices called histories denoted as $\mu$, which leads to $2^m$
possible histories. By taking all histories and assigning choice for
each of them, they get a {\em strategy}. Each player is allowed to
have a limited set of $S$ strategies. She keeps score for each
strategy according to the $m$ last winning choices, i.e.,
distinguishes ``good'' and ``bad'' strategies and chooses the best
strategy with the highest score to make a decision for the next
step~\cite{Challet1997}.

Even this simplest original model of MG shows quite interesting
properties such that the volatility $\sigma$, the standard deviation
of the time series $A(t)$ which counts the number of players
choosing a certain alternative, follows the scaling relation
$\sigma^2 / N = f(2^m / N)$~\cite{Savit1999}. Furthermore, this
scaling function is divided into two phases called symmetric and
asymmetric phases, where the strategy distribution and the system
predictability are different~\cite{Savit1999}. The volatility is
considered as an important quantity because it is the measure of the
system inefficiency, provided that the maximum number of winners is
bounded by $N/2$. This behavior of the volatility has, therefore,
caught much attention from researchers.

Besides such efforts to unveil the structure and properties of the
original MG, some extensions or variations of the original model
are introduced to adopt more realistic aspects or for mathematical
convenience~\cite{Johnson1999,Cavagna1999}. The concept of social
interaction among players in terms of ``imitation,'' superimposed
with the global information, is introduced in
Refs.~\cite{Slanina2000,Eguluz2006}. Later, in
Ref.~\cite{Anghel2003}, players in MG are connected by the Erd{\H
o}s-R{\' e}nyi (ER) random graph~\cite{Erdos1959}.
In their model, an undirected {\em substrate} network {\bf G} of the
ER type is constructed between players before the game. For each
moment of choice, each player first chooses one of two alternatives
according to her own best strategy, but she does not use the choice
immediately. Instead, she compares the points, defined as the number
of victories of each player from the beginning of the game, of her
neighbors connected by the substrate network and follows the choice
of the player with the maximum points. If no neighbor is better than
herself, she makes the final decision by following her own best
strategy. After some time, the {\em leadership structure}, i.e., who
follows whom, appears and is a directed subset {\bf F} of the
underlying substrate network {\bf G}. Anghel {\em et al.}~shows that
despite the Poisson degree distribution of the substrate ER graph,
the incoming degrees of this ``follower'' network {\bf F} are
described by a power-law distribution with a sharp
cut-off~\cite{Anghel2003}, which indicates the {\em scale-free}
structure known to be ubiquitous in the real world network
systems~\cite{Newman2003a,Albert2002,Dorogovtsev2001a}.

In this paper, we generalize the approach of Anghel {\em et al.}~by
varying the types of substrate networks, such as the ER random
graph~\cite{Erdos1959}, regular lattices, small-world
networks~\cite{Watts1998}, and scale-free
networks~\cite{Barabasi1999,KIGoh2001}, since the real social
network structures are known to be much more complicated than the ER
random graph. Furthermore, we observe not only the structure of the
follower networks, but also the volatility $\sigma$.
There have been quite a number of studies that set up ``recently
recognized'' complex network structures in several systems of
interacting components, e.g., the Ising
model~\cite{Pekalski2001,Hong2002,BJKim2001,Lopes2004,DJeong2005,
Meilikhov2005}, the voter
model~\cite{Suchecki2004,Sood2004a,Castellano2005,Chen2005,Lima2004},
the prisoners' dilemma
game~\cite{Nowak1992,BJKim2002,Wu2005a,Zimmermann2005,Santos2005},
the snowdrift game~\cite{Santos2005,LXZhong2006,WXWang2006c}, the
Boolean game~\cite{Zhou2005d}, etc. All these works have provided
opportunities to understand both the systems themselves and the
properties of complex networks better. Similarly, we expect that
this work also reveals some features of both sides.

\section{rules of the game and substrate network structures}
The rules are basically the same as in Ref.~\cite{Anghel2003} as
stated in the previous section, except for using various kinds of
networks as underlying substrate structures. Each player follows her
neighbor with the best performance by imitating the neighbor's
action (choice) at each time step, as well as keeping records of her
own strategies. We set the number of each player's strategies as
$S=2$ and change the memory $m$ to observe the behavior of MG
systems. Similar to other studies on MG, we confirmed that most
qualitative features do not change for other values of $S$.

As undirected substrate networks, we use the ER random
graph~\cite{Erdos1959}, the one-dimensional (1D) regular lattice
where each player is connected to her neighbors two or fewer lattice
spacings away,
the Watts-Strogatz (WS) small-world network~\cite{Watts1998}, the
Barab{\' a}si-Albert (BA) growing scale-free network with the degree
exponent $\gamma = 3$ where the degree distribution $p(k) \sim
k^{-\gamma}$~\cite{Barabasi1999}, and the Goh-Kahng-Kim (GKK) static
scale-free network with $\gamma=2$~\cite{KIGoh2001}. For the
volatility function, especially we focus on the WS case, varying the
rewiring probability $p$, since the volatility function changes its
shape continuously but drastically with the disorder parameter $p$.
In addition, we try some real-world social networks such as a
coauthorship network~\cite{Newman2001a} and bulletin board system
(BBS) networks, composed of BBS users connected by their replies
among each other~\cite{KIGoh2006}, to test the generality of the
results.

\section{volatility function}
The term volatility $\sigma$ in MG is defined as the standard
deviation of the time series $A(t)$, representing the difference
between the number of agents who choose a certain alternative and
the other at time $t$, as follows:
\begin{equation}
\sigma = \sqrt{\langle A^2 (t) \rangle - \langle A(t) \rangle^2 }.
\end{equation}
Here the angular bracket refers to the average over time. If $A(t)$
is a random sequence of two alternatives with equal probabilities,
$\sigma^2 / N = 1/4$, and the situation is the so-called
``coin-toss'' case. Minimizing this volatility corresponds to
maximizing the efficiency of the system, since the amount of
deviation from the average value $N/2$ (by symmetry) is viewed as
the {\em waste} of available resources. In this sense, the
volatility is a reciprocal measure of the system efficiency.

A remarkable result of early studies on the original MG model is
that $\sigma^2 / N$ does not depend independently on $m$ and $N$,
but only on the ratio $\alpha = 2^m / N$~\cite{Savit1999}. This
scaling relation is explained as the effect caused by the relative
size difference in the number of player $N$ and the size of {\em
effective} strategy space $2^m$, related to the distribution and
overlap of strategies among players. For $\alpha \ll 1$, $\sigma^2
\sim 1/\alpha$ and this phase is called a crowded, unpredictable,
and symmetric phase. On the other hand, when $\alpha \gg 1$,
$\sigma^2 \simeq 1/4$ (the system approaches the coin-toss case) and
this phase is called an uncrowded, predictable, and asymmetric
phase. The phase separation has been investigated by analogy with
spin formalism~\cite{Challet1999,Challet2000b,Marsili2001a}.

\begin{figure}
\includegraphics[width=0.45\textwidth]{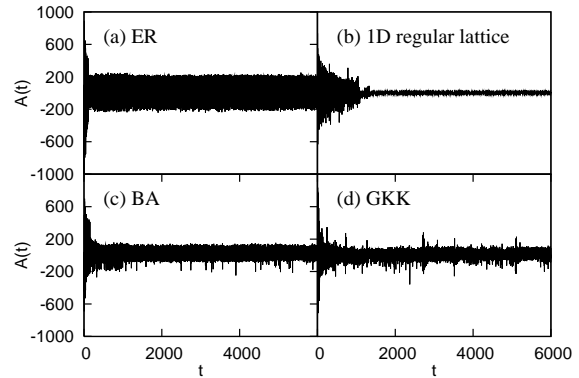}
\caption[0]{Typical time series $A(t)$ with $N=1001$ and $m=2$ for
(a) ER, (b) 1D regular lattice, (c) BA, and (d) GKK substrate
networks.} \label{TimeSeries}
\end{figure}

\begin{figure*}
\includegraphics[width=\textwidth]{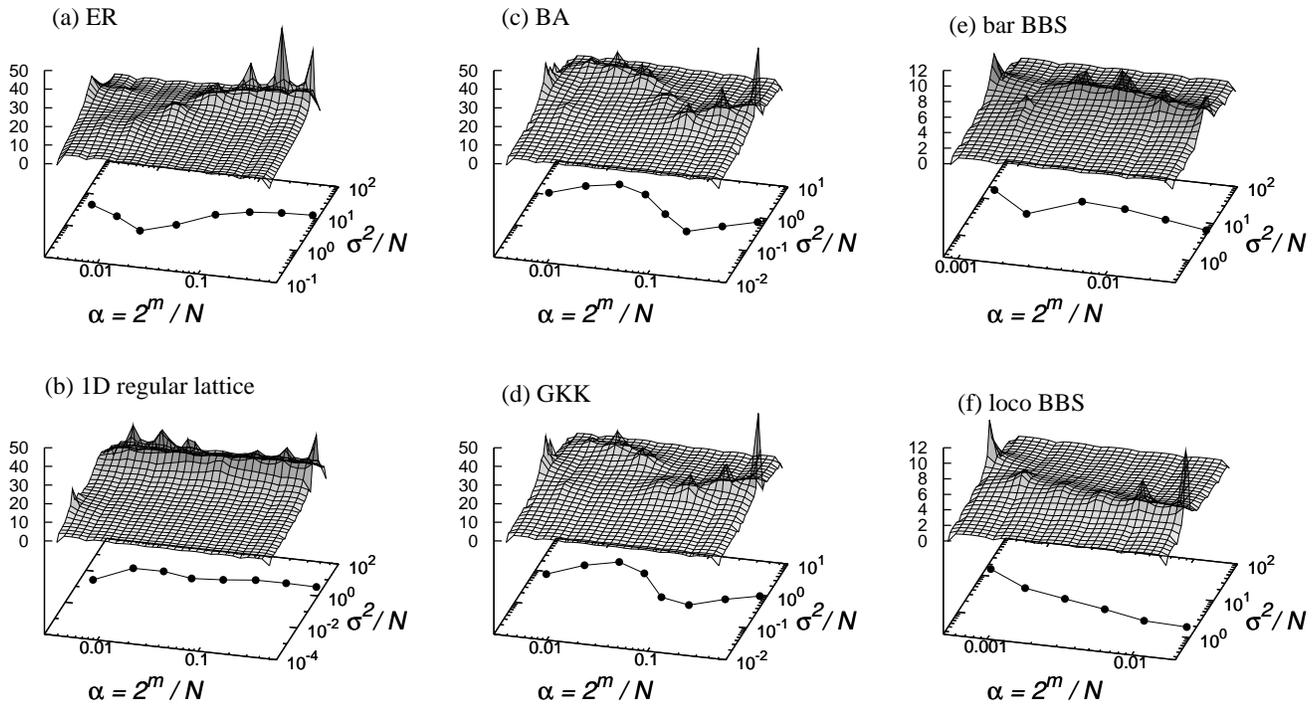}
\caption[0]{Volatility $\sigma^2 / N$ distribution according to the
value $\alpha = 2^m / N$, for (a) ER, (b) 1D regular lattice, (c)
BA, (d) GKK, (e) bar BBS, and (f) loco BBS substrate networks. The
numbers of players are $N = 1001$ for (a)--(d), $N = 4461$ for (e),
and $N = 7410$ for (f). For each value of memory $m$, the results of
500 simulations for (a)--(d) are recorded as a histogram, whereas
(e) and (f) are from 100 realizations. Average values of the
volatility are drawn at the bottom of each graph.} \label{vola_3d}
\end{figure*}

Extensive simulations of MG on various substrate networks show that
in contrast to the case of the original MG where the clear scaling
behavior of volatility is shown, the volatility $\sigma^2 / N$ is
not a function of the single variable $\alpha = 2^m / N$. Moreover,
variance of $\sigma^2 / N$, for each realization, is significantly
large for some cases. Figure~\ref{TimeSeries} shows typical times
series for various substrate networks and the volatility is shown in
Fig.~\ref{vola_3d}. For the ER substrate network, the form of
volatility is similar to the original MG model. Other topologies,
however, totally change the shape of the volatility function against
$\alpha$. For a 1D regular lattice, except for the very small value
of $\alpha$, the volatility is so large that the system can be
considered as inefficient or maladapted. In the case of a very small
value of $\alpha$, on the other hand, the volatility becomes quite
small~\cite{Vola_1D}. [Notice the scale of the volatility axis in
Fig.~\ref{vola_3d}(b)] For $N = 1001$, extremely small volatility
shown in Fig.~\ref{TimeSeries}(b) happens only for $m = 2$, the
smallest value of $m$ used in the simulation, and increasing the
value just to $m = 3$ causes inefficiency. The volatility for $m =
2$ in this case (1D regular lattice) is observed as the smallest
volatility among all the simulation results. Therefore, from the
perspective of the system efficiency defined as the reciprocal of
the volatility, a 1D regular lattice as substrate and a very small
value of $\alpha$ are the best ingredients for an efficient system.

For BA and GKK substrate networks, the shapes of volatility against
$\alpha$ quite resemble each other, in spite of the difference in
the creation mechanism~\cite{Barabasi1999,KIGoh2001} and the degree
exponent $\gamma$. [Compare Figs.~\ref{vola_3d}(c) and
\ref{vola_3d}(d)] The peculiar part, compared with the original MG
or MG on ER substrate network, is the increasing behavior of
volatility for small $\alpha$. Since MG stems from economics where
social networks among people are important, we also use BBS networks
in a university~\cite{KIGoh2006} as examples of real social networks
for the simulation. The two BBS networks have their own
characteristic shapes of the volatility as shown in
Figs.~\ref{vola_3d}(e) and \ref{vola_3d}(f). From the simulation
result, it is clear that the topology of substrate networks
significantly influences the dependence of system volatility on the
parameter $\alpha$. Especially, there is a drastic difference
between the ER random graph and 1D regular lattice because the
randomness or disorderedness is extremely different, although both
of these two networks have homogeneous degree distributions. The
scale-free networks with heterogeneous degree distributions seem to
be located somewhere in between those two extremes. The fact that
the underlying network topology effects on the ability to
communicate and the information transfer has been shown
recently~\cite{Sneppen2004} and the volatility function of MG on
networks here is another good example.

\begin{figure}
\includegraphics[width=0.45\textwidth]{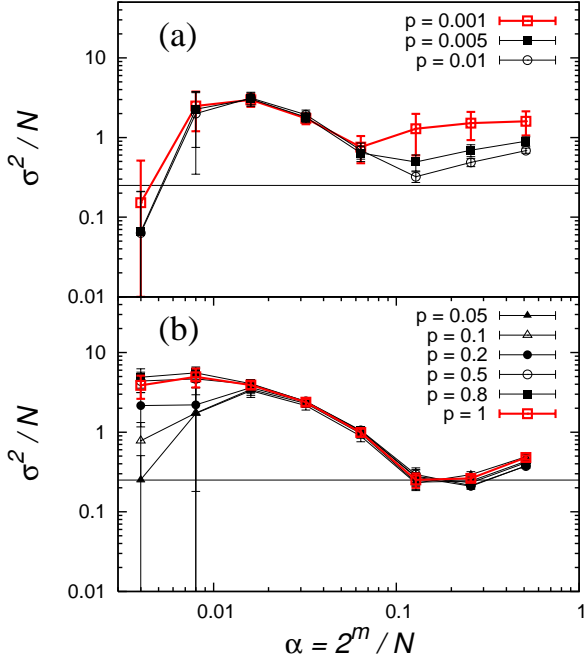}
\caption[0]{(Color online) Volatility $\sigma^2 / N$ according to
the value $\alpha = 2^m / N$ for WS small-world network with (a)
small, and (b) large values of rewiring probability $p$. The number
of players is $N = 1001$, and each player is connected to her
neighbors two or fewer lattice spacings away. For each value of
memory $m$, the results of 500 simulations are averaged. The solid
lines are guides to the eyes, and the horizontal line ($\sigma^2 / N
= 0.25$) indicates the coin-toss case. Each (red) line corresponds
to a limiting case of $p$, i.e., the 1D regular lattice in (a) and
the ER random graph in (b), respectively.} \label{vola_sw}
\end{figure}

For the more systematic approach, we use the WS small-world
network~\cite{Watts1998}. One convenient aspect about the WS network
is that we can control the disorderedness of the structure by
changing the rewiring probability $p$. Figure~\ref{vola_sw} shows
the result of volatility from MG on WS networks with different
values of $p$. By definition of the rewiring probability $p$, the
case of $p \to 0$ ($p \to 1$) corresponds to the 1D regular lattice
(ER random network), respectively. As one can imagine, increasing
$p$ from $0$ to $1$ (adding disorderedness to the substrate network)
continuously changes the form of volatility from the 1D regular
lattice case to the ER random network case. Figure~\ref{vola_sw}(a)
and (b), furthermore, show a certain pattern of change depending on
the value of $p$. For small values of $p$ (Fig.~\ref{vola_sw}(a)),
the volatility for large values of $\alpha$ changes rapidly as $p$
increases. On the other hand, for large values of $p$
(Fig.~\ref{vola_sw}(b)), the volatility for small values of $\alpha$
changes rapidly as $p$ increases, while that for large $\alpha$ is
almost fixed.

The region of small values of $\alpha = 2^m / N$ is traditionally
called the symmetric or crowded phase in the original MG, which
means that the number of players $N$ is much larger than the
effective size of the strategy space, so that many players come to
use the same strategies. Large values of $\alpha$ correspond to the
asymmetric, anticrowded, or {\em information-rich} phase, because
the lack of overusage of strategies causes inequality of strategies'
scores in the whole system, which leads to the possibility of
exploiting the information and predicting. According to the
properties of WS small-world networks, the average path length of
the network drops dramatically as $p$ increases, even for small
values of $p$, while the clustering coefficient finally drops for
large values of $p$. Therefore, we can conclude that for small
values of $p$, a radically decreasing average path length, possibly
related to the efficiency of the information transfer, affects the
information-rich phase, i.e., the anticrowded phase for the large
value of $\alpha$. For large values of $p$, a sharply dropping value
of clustering coefficient seems to affect the crowded effect for the
crowded phase with a small value of $\alpha$. From this result, we
can see that the disorderedness of the substrate networks clearly
influences the system behavior, in regard to the volatility.

\begin{figure}
\includegraphics[width=0.45\textwidth]{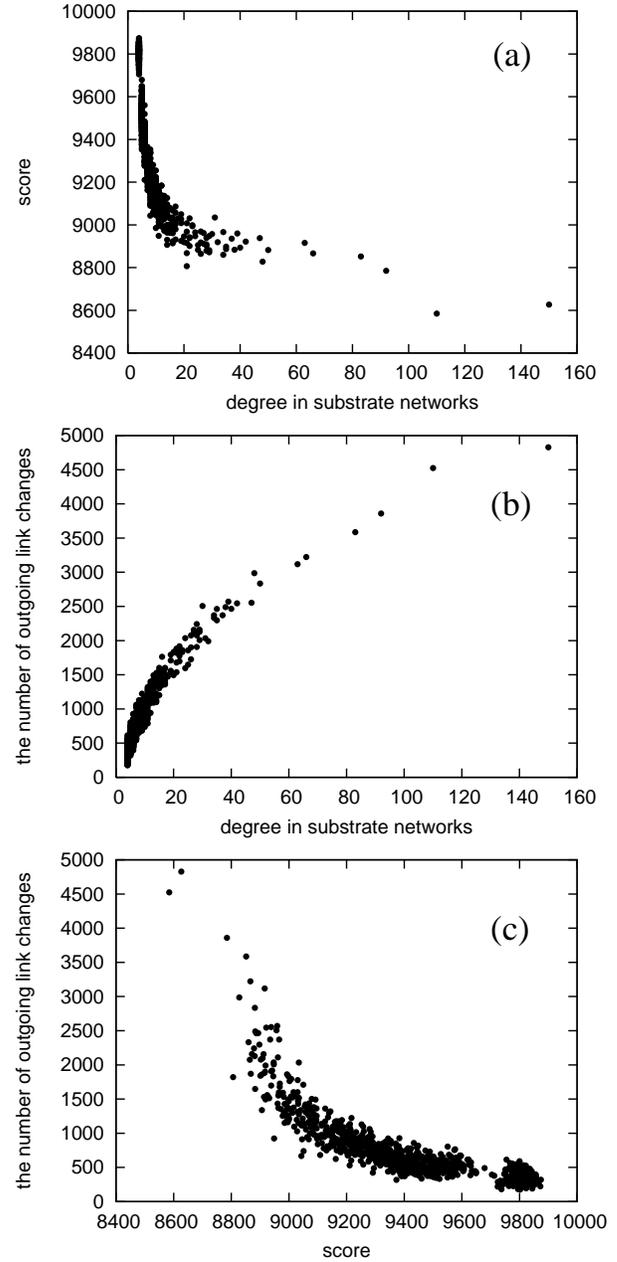}
\caption[0]{The interrelationship among the average scores of
players, their degrees in the substrate network, and the average
numbers of outgoing link changes per game. Each data point
represents the values corresponding to each player averaged over 500
games, each of which is composed of 20000 steps. ($m = 2$) The
substrate network is the BA model with 1001 nodes.}
\label{DegreeScore}
\end{figure}

It is worthwhile to note that for most regions of parameters and
types of substrate networks, the volatility is {\em larger} than
that in the original MG and even the random coin-toss case. In other
words, the system with networks performs {\em less} effectively in
general. The reason why the original MG system cannot reach a Nash
equilibrium, i.e., $\sigma \sim \mathcal{O}(1)$, is that each player
does not consider her own ``market impact'' on the game and causes
the herding behavior~\cite{Marsili2001}. Adding substrate networks,
which enable players to interact with one another, even more
accelerates this herding effect thanks to the locally common
references and leads to the inefficiency. Basically, each player
plays the game without any {\em centralized} control and this sort
of inefficiency called ``price of anarchy,'' due to such
decentralization, has gotten attention recently~\cite{POA}. An onset
of bubble phenomena by rumors from peers is not unusual in economy,
which is a good example of the social network effect.

Figure~\ref{DegreeScore}(a) shows that a ``hub'' with many
connections in the first place is likely to gain a lower score than
other players, due to the herding behavior of the players following
the hub. A hub tends to be more likely to change its outgoing link
in the follower network during a game, as shown in
Fig.~\ref{DegreeScore}(b). We can easily see that a player's score
and the frequency of the outgoing link change are closely related as
well, from Fig.~\ref{DegreeScore}(c). In sum, hubs are likely to
lose due to their followers, which leads to the frequent changes of
their outgoing links in the follower network. For many other
networks and parameters, this pattern is observed in common. Players
with large degrees, therefore, are not able to take advantage of the
situation. Oppositely, they suffer from the lowered chance to win
because of their followers who will make the same choice with them,
which of course the hub players do not want, considering the very
definition of the minority rule.

\begin{figure}
\includegraphics[width=0.45\textwidth]{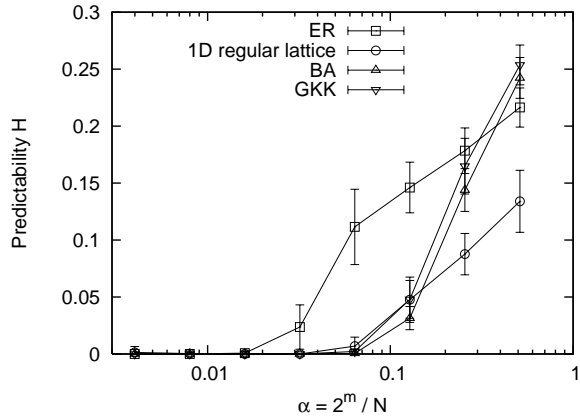}
\caption[0]{Predictability $H$ according to the value $\alpha = 2^m
/ N$, for each substrate network whose size $N = 1001$. Each point
and error bar are drawn values averaged over 100 games, each of
which is composed of 20000 steps.} \label{Predictability}
\end{figure}

Besides the volatility, the predictability $H$ is also an important
quantity in the MG system and characterizes the phase transition.
The predictability is defined as~\cite{ChalletBook,CoolenBook},
\begin{equation}
H = \frac{1}{2^m} \sum_{\mu} \langle A | \mu \rangle^2, \label{H}
\end{equation}
where $\mu$ is the $m$-bit history, the summation over $\mu$ is for
all the possible $2^m$ values of $\mu$, and $\langle A | \mu
\rangle$ is the temporal average value of the outcomes of the game
for each history $\mu$, composed of the last $m$ bits of winning
choice. The predictability, therefore, measures the
``deterministic'' tendency of the game for each given history. The
larger values of the predictability implies that the outcome of the
game at each time is more predictable and each player tends to stick
to a single strategy, which gives the analogy of the ordered spin
state in the spin-glass theory.

In the original MG model, the predictability is in the vicinity of
$0$ at the symmetric phase (corresponding to small $\alpha = 2^m /
N$ values) and starts to increase with the onset of the asymmetric
phase as $\alpha$ increases. This characteristic of $H$ according to
$\alpha$ is the reason why the symmetric phase is called the
unpredictable phase and the asymmetric phase is called the
predictable phase. Figure~\ref{Predictability} shows the
predictability values according to $\alpha$ for our MG model with
various substrate network structures. Like the original MG model,
the predictability starts to increase from a certain threshold value
of $\alpha$, below which the predictability is essentially $0$. The
threshold values, however, depend on substrate network topologies.
Comparing Figs.~\ref{vola_3d}(a)--\ref{vola_3d}(d) and
\ref{Predictability}, we can see that, for MG on ER random networks,
the volatility reaches its minimum value and the predictability
begins to increase for smaller value of $\alpha$ than for the game
on the other networks.

\section{structures of the follower networks}

\begin{figure}
\includegraphics[width=0.45\textwidth]{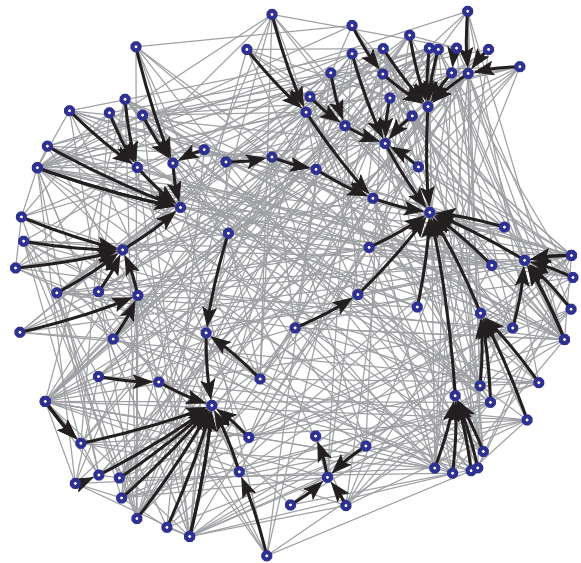}
\caption[0]{(Color online) Snapshot of typical structure of follower
network, where the ER random graph is used as substrate in this
case. Thin gray lines stand for the substrate network and thick
black arrows for the follower network.} \label{Pajek}
\end{figure}

\begin{figure}
\includegraphics[width=0.45\textwidth]{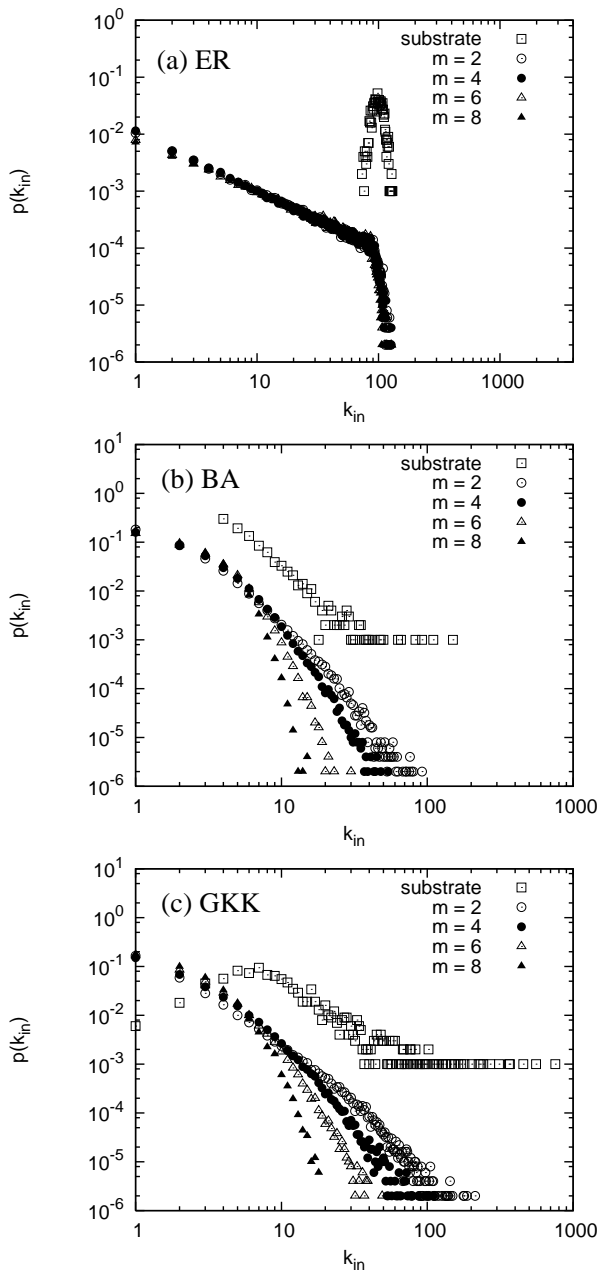}
\caption[0]{Incoming degree distribution of the follower networks
for $N = 1001$ whose substrate networks are (a) the ER random graph
with connecting probability $p=0.1$~\cite{Erdos1959}, (b) the BA
model with $m_0 = m = 4$~\cite{Barabasi1999}, and (c) the GKK model
with $\gamma = 2$ and $\textrm{average number of degrees} =
10$~\cite{KIGoh2001}. Empty squares ($\square$) correspond to the
degrees of substrate networks. The degree distribution of each
follower network is drawn from 500 realizations of MG.}
\label{follower_degree}
\end{figure}

\begin{figure}
\includegraphics[width=0.45\textwidth]{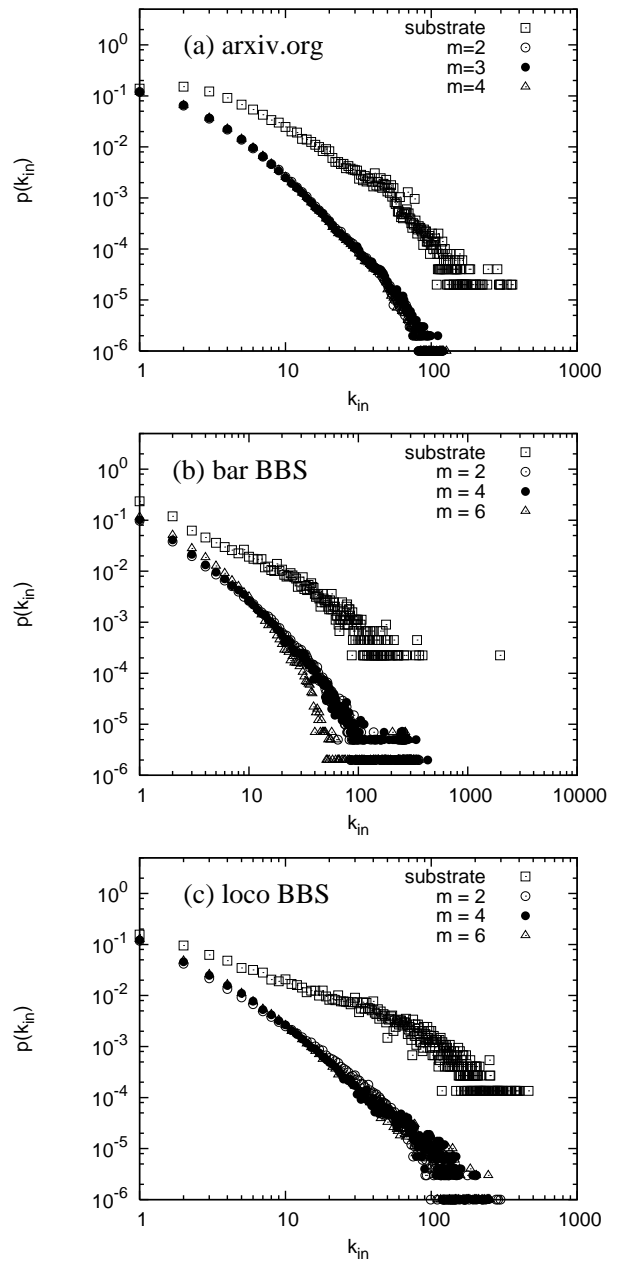}
\caption[0]{Incoming degree distribution of the follower networks
for whose substrate networks are (a) the e-print archive
coauthorship network with 49983 nodes~\cite{Newman2001a}, (b) the
bar BBS network with 4461 nodes~\cite{KIGoh2006}, and (c) the loco
BBS network with 7410 nodes~\cite{KIGoh2006}. Empty squares
($\square$) correspond to the degrees of substrate networks. The
degree distribution of each follower network is drawn from 100
realizations of MG.} \label{social_follower}
\end{figure}

After the transient period, the follower network structure reaches
its steady state and only a small fraction of nodes change their
outgoing link at each time step (less than ten percent even for the
worst case). In addition, even though some nodes change their
outgoing link, the characteristic features such as the degree
distribution do not change. Therefore, we analyze the structure of
the follower networks for each substrate network topology.

Figure~\ref{Pajek} shows a typical structure of follower networks
for the case of the ER substrate network~\cite{PajekProgram}. If
node $a$ follows node $b$, the direction of link $a \to b$ is
established. Hubs with large incoming degree (in-degree) are easily
noticed. In-degree distribution of follower networks shown in
Fig.~\ref{follower_degree} implies the scale-free structure of
follower networks for (a) ER, (b) BA, and (c) GKK substrate
networks. We also observed the same results of the fat-tailed
in-degree distributions for the e-print archive coauthorship
network~\cite{Newman2001a} and BBS networks~\cite{KIGoh2006}, as
shown in Fig.~\ref{social_follower}. Cutoffs in
Fig.~\ref{follower_degree}(a), also stated in
Ref.~\cite{Anghel2003}, are due to the fact that the number of links
for each node is bounded by the Poisson degree distribution of the
substrate network. In this respect, the scale-free structure is
conserved for MG. In addition, from Figs.~\ref{follower_degree} and
\ref{social_follower}, we can see that the larger the memory size
$m$ is, the narrower the distribution gets. In other words, larger
memory sizes of individual players tend to lower the possibility of
the emergence of large hubs with many connections. Interestingly,
this phenomenon can be interpreted in that ``smarter'' players
effectively prevent the appearance of ``dictators'' with big power.

\section{discussion and conclusions}

Even though both MG and complex networks have been heavily studied
to understand the structure and dynamics of complex systems, the
interaction among players in the MG system via complex networks has
not drawn much attention so far. By numerically studying this
subject, we have shown that the topology of underlying substrate
networks significantly influences the well-known properties of MG.
We also demonstrate the emergence of scale-free coordination
structures, as suggested in earlier studies.

Volatility, traditionally considered as the reciprocal measure of
the system efficiency, of MG on various networks has quite different
structures from the original MG, depending on the topology of the
substrate. Especially, MG on 1D regular lattice with very small
memory size seems to exploit the system effectively as shown in
Fig.~\ref{TimeSeries}(b), although the information transfer is known
not to be efficient in regular lattices. In contrast, in most cases
the performance of systems is worse than the original MG or the
random choice, due to the enhanced herding behavior provided by
interactions via substrate networks. Disorderedness of the substrate
networks, reflected by the rewiring probability $p$ of the WS
small-world network, affects the form of the volatility function in
different domains, depending on the degree of disorderedness.

MG is famous for its simple rules, and at the same time for its
complex structure including many counterintuitive aspects. We have
shown that the properties of substrate interaction topology, such as
the degree distribution and randomness, play a great role in the
system. Still, there are many open questions about the qualitative
and quantitative properties of this system. The original MG has been
investigated not only by numerical simulation, but also by analytic
approaches which have accelerated the understanding of this complex
game. The analytic approach of MG with interactions among players
has also started~\cite{Anghel2003}, and we hope that more research
on this subject of the emergence of complex behavior, including MG
with interactions, will be continued in the future.

\begin{acknowledgments}
This work was supported by KOSEF through Grant No.
R01-2005-000-1112-0, and H.J. acknowledges financial support from
KRF (MOEHRD) through Grant No. R14-2002-059-01000-0.
\end{acknowledgments}

\begin{thebibliography}{99}
\bibitem{Vola_1D} But even in this case, a small number of realizations show
large volatility. However, the odds are rare.
\end{thebibliography}
\end{document}